\documentclass[useAMS,usenatbib]{mn2e}

\usepackage[dvips]{graphicx}
\usepackage{lineno}
\usepackage{amsmath}
\usepackage{hyperref}
\usepackage{breakurl}
\title[VERITAS and Multiwavelength Observations of the BL Lacertae Object 1ES 1741+196]{VERITAS and Multiwavelength Observations of the BL Lacertae Object 1ES 1741+196}

\author[VERITAS Collaboration]
{A.~U.~Abeysekara$^{1}$,
S.~Archambault$^{2}$, 
A.~Archer$^{3}$, 
W.~Benbow$^{4}$, 
R.~Bird$^{5}$, 
J.~Biteau$^{6}$,\newauthor
M.~Buchovecky$^{7}$, 
J.~H.~Buckley$^{3}$, 
V.~Bugaev$^{3}$, 
K.~Byrum$^{8}$, 
J.~V~Cardenzana$^{9}$,\newauthor
M.~Cerruti$^{4}$, 
X.~Chen$^{10,11}$, 
J.~L.~Christiansen$^{12}$\thanks{jlchrist@calpoly.edu}, 
L.~Ciupik$^{13}$, 
M.~P.~Connolly$^{14}$,\newauthor
W.~Cui$^{15}$,
H.~J.~Dickinson$^{9}$, 
J.~Dumm$^{16}$, 
J.~D.~Eisch$^{9}$, 
M.~Errando$^{17}$, 
A.~Falcone$^{18}$,\newauthor  
Q.~Feng$^{15}$,
J.~P.~Finley$^{15}$, 
H.~Fleischhack$^{11}$, 
A.~Flinders$^{1}$, 
P.~Fortin$^{4}$, 
L.~Fortson$^{16}$,\newauthor 
A.~Furniss$^{19}$,  
G.~H.~Gillanders$^{14}$, 
S.~Griffin$^{2}$, 
J.~Grube$^{13}$, 
G.~Gyuk$^{13}$, 
M.~Huetten$^{11}$,\newauthor 
D.~Hanna$^{2}$, 
J.~Holder$^{20}$, 
T.~B.~Humensky$^{21}$, 
C.~A.~Johnson$^{6}$, 
P.~Kaaret$^{22}$, 
P.~Kar$^{1}$,\newauthor 
N.~Kelley-Hoskins$^{11}$, 
M.~Kertzman$^{23}$, 
D.~Kieda$^{1}$, 
M.~Krause$^{11}$, 
F.~Krennrich$^{9}$,\newauthor 
M.~J.~Lang$^{14}$,
G.~Maier$^{11}$, 
S.~McArthur$^{15}$, 
A.~McCann$^{2}$, 
K.~Meagher$^{24}$, 
P.~Moriarty$^{14}$,\newauthor  
R.~Mukherjee$^{17}$, 
D.~Nieto$^{21}$, 
S.~O'Brien$^{5}$, 
A.~O'Faol\'{a}in de Bhr\'{o}ithe$^{11}$, 
R.~A.~Ong$^{7}$,\newauthor  
A.~N.~Otte$^{24}$, 
N.~Park$^{25}$, 
V.~Pelassa$^{4}$, 
A.~Petrashyk$^{21}$,
D.~Petry$^{26}$, 
M.~Pohl$^{10,11}$,\newauthor  
A.~Popkow$^{7}$, 
E.~Pueschel$^{5}$\thanks{elisa.pueschel@ucd.ie}, 
J.~Quinn$^{5}$, 
K.~Ragan$^{2}$, 
G.~Ratliff$^{13}$, 
L.~C.~Reyes$^{12}$,\newauthor 
P.~T.~Reynolds$^{27}$,
K.~Reynolds$^{12}$, 
G.~T.~Richards$^{24}$, 
E.~Roache$^{4}$, 
C.~Rulten$^{16}$,\newauthor 
M.~Santander$^{17}$, 
G.~H.~Sembroski$^{15}$, 
K.~Shahinyan$^{16}$, 
A.~W.~Smith$^{28}$, 
D.~Staszak$^{2}$,\newauthor 
I.~Telezhinsky$^{10,11}$, 
J.~V.~Tucci$^{15}$, 
J.~Tyler$^{2}$, 
S.~Vincent$^{11}$, 
S.~P.~Wakely$^{25}$, 
O.~M.~Weiner$^{21}$,\newauthor 
A.~Weinstein$^{9}$, 
A.~Wilhelm$^{10,11}$, 
D.~A.~Williams$^{6}$, 
B.~Zitzer$^{8}$\\
$^{1}$Department of Physics and Astronomy, University of Utah, Salt Lake City, UT 84112, USA\\
$^{2}$Physics Department, McGill University, Montreal, QC H3A 2T8, Canada\\
$^{3}$Department of Physics, Washington University, St. Louis, MO 63130, USA\\
$^{4}$Fred Lawrence Whipple Observatory, Harvard-Smithsonian Center for Astrophysics, Amado, AZ 85645, USA\\
$^{5}$School of Physics, University College Dublin, Belfield, Dublin 4, Ireland\\
$^{6}$Santa Cruz Institute for Particle Physics and Department of Physics, University of California, Santa Cruz, CA 95064, USA\\
$^{7}$Department of Physics and Astronomy, University of California, Los Angeles, CA 90095, USA\\
$^{8}$Argonne National Laboratory, 9700 S. Cass Avenue, Argonne, IL 60439, USA\\
$^{9}$Department of Physics and Astronomy, Iowa State University, Ames, IA 50011, USA\\
$^{10}$Institute of Physics and Astronomy, University of Potsdam, 14476 Potsdam-Golm, Germany\\
$^{11}$DESY, Platanenallee 6, 15738 Zeuthen, Germany\\
$^{12}$Physics Department, California Polytechnic State University, San Luis Obispo, CA 94307, USA\\
$^{13}$Astronomy Department, Adler Planetarium and Astronomy Museum, Chicago, IL 60605, USA\\
$^{14}$School of Physics, National University of Ireland Galway, University Road, Galway, Ireland\\
$^{15}$Department of Physics and Astronomy, Purdue University, West Lafayette, IN 47907, USA\\
$^{16}$School of Physics and Astronomy, University of Minnesota, Minneapolis, MN 55455, USA\\
$^{17}$Department of Physics and Astronomy, Barnard College, Columbia University, NY 10027, USA\\
$^{18}$Department of Astronomy and Astrophysics, 525 Davey Lab, Pennsylvania State University, University Park, PA 16802, USA\\
$^{19}$Department of Physics, California State University - East Bay, Hayward, CA 94542, USA\\
$^{20}$Department of Physics and Astronomy and the Bartol Research Institute, University of Delaware, Newark, DE 19716, USA\\
$^{21}$Physics Department, Columbia University, New York, NY 10027, USA\\
$^{22}$Department of Physics and Astronomy, University of Iowa, Van Allen Hall, Iowa City, IA 52242, USA\\
$^{23}$Department of Physics and Astronomy, DePauw University, Greencastle, IN 46135-0037, USA\\
$^{24}$School of Physics and Center for Relativistic Astrophysics, Georgia Institute of Technology, 837 State Street NW, Atlanta, GA 30332-0430\\
$^{25}$Enrico Fermi Institute, University of Chicago, Chicago, IL 60637, USA\\
$^{26}$ALMA Regional Centre, ESO, Karl-Schwarzschild-Str. 2, D-8
5748 Garching, Germany\\
$^{27}$Department of Physical Sciences, Cork Institute of Technology, Bishopstown, Cork, Ireland\\
$^{28}$University of Maryland, College Park / NASA GSFC, College Park, MD 20742, USA\\
}

\begin{document}

\date{Submitted \today}

\pagerange{\pageref{firstpage}--\pageref{lastpage}} \pubyear{2015}

\maketitle

\label{firstpage}

\begin{abstract}
We present results from multiwavelength observations of the BL Lacertae object 1ES 1741+196, including results in the very-high-energy $\gamma$-ray regime using the Very Energetic Radiation Imaging Telescope Array System (VERITAS).  The VERITAS time-averaged spectrum, measured above 180 GeV, is well-modelled by a power law with a spectral index of $2.7\pm0.7_{\mathrm{stat}}\pm0.2_{\mathrm{syst}}$. The integral flux above 180 GeV is $(3.9\pm0.8_{\mathrm{stat}}\pm1.0_{\mathrm{syst}})\times 10^{-8}$ m$^{-2}$ s$^{-1}$, corresponding to 1.6\% of the Crab Nebula flux on average. The multiwavelength spectral energy distribution of the source suggests that 1ES 1741+196 is an extreme-high-frequency-peaked BL Lacertae object. The observations analysed in this paper extend over a period of six years, during which time no strong flares were observed in any band. This analysis is therefore one of the few characterizations of a blazar in a non-flaring state.
\end{abstract}


\section{Introduction}
BL Lacertae (BL Lac) objects constitute the majority of the population of extragalactic objects detected in the very-high-energy (VHE; $>$100~GeV) $\gamma$-ray band. They are  characterized by a featureless optical spectrum and a double-humped spectral energy distribution (SED) with a lower-energy synchrotron peak and a higher-energy peak often attributed to inverse-Compton scattering. BL Lacs are further classified based on the location of their synchrotron peak as low-, intermediate-, or high-frequency-peaked (LBL, IBL, HBL)~\citep{Padovani1995}. The synchrotron peak of an HBL lies in the ultraviolet or X--ray range, whereas that of an LBL lies in the optical or infrared range. Due to its synchrotron peak above 1~keV, the $\gamma$-ray source 1ES 1741+196 is classified as an HBL~\citep{Nieppola}. The spectral energy distribution (SED) presented in this paper points to further classification as an extreme-HBL, as will be discussed in Section~\ref{discussion}.

1ES 1741+196 was initially identified as a BL Lac by the \textit{Einstein} ``Slew Survey"~\citep{Perlman}. Data from this X-ray survey were used in conjunction with radio data to identify a number of BL Lacs by searching for radio and X-ray loud objects without strongly-identified emission lines. Further radio studies by \cite{Rector} clearly demonstrated the presence of a well-collimated jet, based on a VLBA map at 4.964 GHz. The source, 1ES 1741+196, was included in a multiwavelength (MWL), multi-blazar study by~\cite{Giommi} that utilized data from \textit{Planck} (microwave), the ROSAT All-Sky Survey Bright Source Catalog (soft X-ray), and \textit{Fermi}--LAT (MeV-GeV $\gamma$-ray). The SED in their paper focuses on simultaneous \textit{Planck} and \textit{Fermi}--LAT observations that result in flux upper limits. 

In the optical range, 1ES 1741+196 was extensively studied by~\cite{Heidt}. Imaging studies of the host galaxy of 1ES 1741+196 and its two companion galaxies were performed using the Nordic Optical Telescope, and spectroscopic studies were performed at the Calar Alto 3.5m telescope. Optical imaging suggested that the host galaxy of 1ES 1741+196 is an elliptical galaxy flattened by tidal interactions with its two companion galaxies. A tidal tail was also observed between the two companion galaxies, indicating further interaction within the triplet. Measurements of the absorption spectrum indicated a redshift of $z$=0.084$\pm$0.001, consistent with~\cite{Perlman}.

The source was detected by \textit{Fermi}--LAT in the high-energy (HE; $>$10~MeV) $\gamma$-ray band, and was included in the 1FGL catalog under the name 1FGL J1744.2+1934~\citep{1FGL}. It was more recently included in the 3LAC catalog as HBL 3FGL J1743.9+1934~\citep{3LAC}. The measured spectral index of $\Gamma$=1.777$\pm0.108_{\mathrm{stat}}$~\citep{3LAC} indicates a source with a hard intrinsic energy spectrum that may extend well into the TeV range. This property, together with the distance to the source, makes 1ES 1741+196 an attractive candidate for studies of the extragalactic background light (EBL), as will be discussed in Section~\ref{discussion}. Measurements of the energy spectrum above 100 GeV may be used to constrain EBL models, particularly in combination with spectral measurements of other hard-spectrum VHE blazars.

Gamma-ray emission from the BL Lac 1ES 1741+196 was first detected at VHE energies by the MAGIC collaboration~\citep{MAGICdetection}. Its discovery as a VHE-emitter was announced in 2011 following 60 hours of observation. MAGIC reported that the integral flux above 250~GeV corresponds to 0.8\% of the Crab Nebula flux.  This makes 1ES 1741+196 one of the weakest BL Lac objects detected by ground-based instruments to date. 

In this paper we present a multiwavelength SED for 1ES 1741+196.  This includes the first-measured VHE photon spectrum of this source, based on $\sim$30 hours of VERITAS observation, spread over six years. We performed analysis covering the optical to the VHE bands, using data from Super-LOTIS, \textit{Swift}, \textit{Fermi}-LAT and VERITAS. We compare the SED to a single-zone synchrotron-self-Compton (SSC) model with reasonable parameters. Additionally, we assess the variability of the source. 

This paper is organized as follows: VHE observations with the VERITAS detector are described in Section~\ref{VERITAS}. Section~\ref{Fermi} describes observations in the HE $\gamma$-ray band with \textit{Fermi}--LAT (both contemporaneous with VERITAS observations and over the full \textit{Fermi}--LAT data set). X-ray observations with \textit{Swift} are discussed in Section~\ref{Swift}. The multiwavelength SED modelling is presented in Section~\ref{SED}, followed by our discussion and conclusions in Sections~\ref{discussion} and~\ref{conclusions}. 

\section{Observations}\label{observations}

\subsection{VERITAS}\label{VERITAS}
The Very Energetic Radiation Imaging Telescope Array System (VERITAS) is an array of imaging atmospheric Cherenkov telescopes located at Fred Lawrence Whipple Observatory in southern Arizona (31$^{\circ}$40$^{\prime}$N, 110$^{\circ}$57$^{\prime}$W). VERITAS consists of four 12-m diameter telescopes of Davies-Cotton design. Each telescope is instrumented with a 499 photomultiplier tube (PMT) camera. Each PMT has a field of view of 0.15$^{\circ}$, combining to a camera field of view of 3.5$^{\circ}$. The instrument is sensitive to VHE $\gamma$-rays between 85 GeV and 30 TeV. The energy resolution is about 20\% and the angular resolution is about 0.1$^{\circ}$ at 1 TeV~\citep{VERITASdetector, VERITASdetector2}.

VERITAS observations of 1ES 1741+196 were made between 19 April 2009 (MJD 54940) and 26 June 2014 (MJD 56834), resulting in a total dead-time-corrected exposure of $\sim$30 hours after removing data affected by poor weather or technical problems. The data-taking period encompasses two major changes in the configuration of the VERITAS array: in summer 2009, one of the telescopes was relocated to create a more symmetric array, and in summer 2012, the cameras were upgraded, enabling the array to operate at a lower energy threshold. All data used in this analysis were collected in ``wobble" mode, meaning that the telescopes are pointed some small angle (0.5$^{\circ}$, in the case of this data set) away from the source location, allowing the signal and background regions to lie within the same field of view and to be observed simultaneously~\citep{Fomin}. The mean zenith angle of the observations was 18$^{\circ}$.

The data were reduced by fitting the shower images in individual cameras with two-dimensional Gaussian functions~\citep{Christiansen}. The selection for separating $\gamma$-ray images from background cosmic-ray images was optimised using Crab Nebula data scaled to 1\% of the standard signal strength. The $\gamma$-ray selection requirements result in an average energy threshold of about 180 GeV for the conditions under which the source was observed.

A circular area centred on the source position was used to define the signal region. The ring-background method was used to define the background region~\citep{Berge}. The event counts in the signal and background regions were used, together with the ratio of the areas of the two regions ($\alpha$=0.044), to calculate the source detection significance according to the prescription of~\cite{LiAndMa}. The observed source significance is 5.9$\sigma$, with a $\gamma$-ray rate of 0.073~$\pm$~0.013~$\gamma$/minute and background cosmic-ray rate of 0.238~$\pm$~0.002~events/minute. 

The source position is determined using Gaussian fits in RA and DEC to a map of the excess events ($N_{\mathrm{on}}$ - $N_{\mathrm{off}}$). The centroid is located at J2000 RA $17^{h} 44^{m} 1.2^{s} \pm 2.4^{s}_{\mathrm{stat}}$ and DEC $+19^{\circ} 32\arcmin 47\arcsec \pm 47\arcsec_{\mathrm{stat}}$. A systematic uncertainty of $\sim$25$\arcsec$ is present in addition to the statistical uncertainty. The uncertainty comes largely from the accuracy of the calibration of the VERITAS pointing system, which corrects for the bending of the telescopes' optical support structures~\citep{Griffiths}. Based on the reconstructed source position, we assign the source the name VER J1744+195. The reconstructed source position is consistent with the nominal position measured by~\cite{VLBAsourcepos}. 

The distribution of the angular distance between the reconstructed arrival direction of the shower and the nominal source position is shown in Figure~\ref{fig:theta2}. In this case, and the spectral analysis that follows, the background was determined using the reflected-region method~\citep{Berge}. The source extension is consistent with emission by a point-like source.

\setcounter{figure}{0}
\begin{figure}
	\includegraphics[width=90mm]{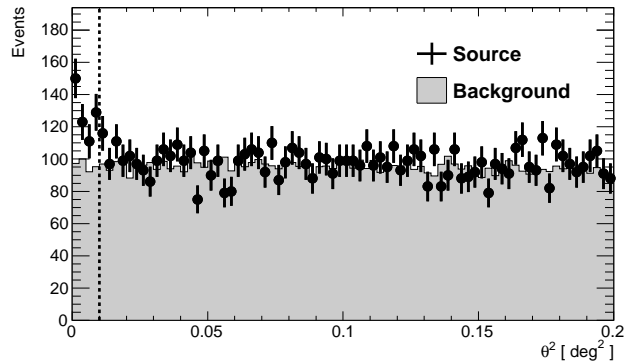}
	\caption{The distribution of the angular distance between the reconstructed arrival direction of the shower and the nominal source position for events reconstructed by VERITAS. The area to the left of the dashed line is the signal region.}
	\label{fig:theta2}
\end{figure}

\setcounter{figure}{1}
\begin{figure}
	\includegraphics[width=84mm]{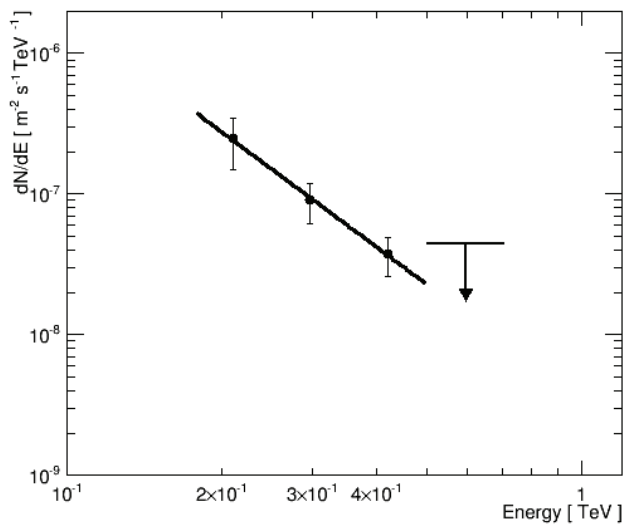}	
	\caption{VERITAS photon spectrum. The fitted spectrum is consistent with a power law.}
	\label{fig:spec}
\end{figure}

The VHE photon spectrum is consistent with a power law, as shown in Fig.~\ref{fig:spec}.  The resulting $\chi^{2}/NDOF$ of 0.03/1 indicates a fit probability of 86\%. The spectrum is characterized by
\begin{multline}
\frac{dN}{dE}  = (9.2 \pm 1.8) \times 10^{-8} \\ \bigg( \frac{E}{0.3~TeV} \bigg)^{-2.7 \pm 0.7} [m^{-2}s^{-1}TeV^{-1}] 
\end{multline}
where the quoted errors are statistical. Based on extensive studies~\citep{Madhavan}, the systematic uncertainty was found to be 25\% on the flux normalization and 0.2 on the spectral index. The spectral points are tabulated in Table~\ref{tab:spectralpts}. The measured integral flux above 180 GeV of $(3.9\pm0.8_{\mathrm{stat}}\pm1.0_{\mathrm{syst}})\times 10^{-8}$ m$^{-2}$ s$^{-1}$ or 1.6 $\pm$ 0.5\% of the Crab Nebula flux where the statistical and systematic uncertainty have been added in quadrature.  This is marginally higher than but still consistent with the MAGIC value of 0.8\% of the Crab Nebula flux~\citep{MAGICdetection}. The calculation of Crab Nebula flux uses the measurement of ~\cite{Albert2008} for a straightforward comparison.

\begin{figure}
\begin{center}
\includegraphics[width=250pt]{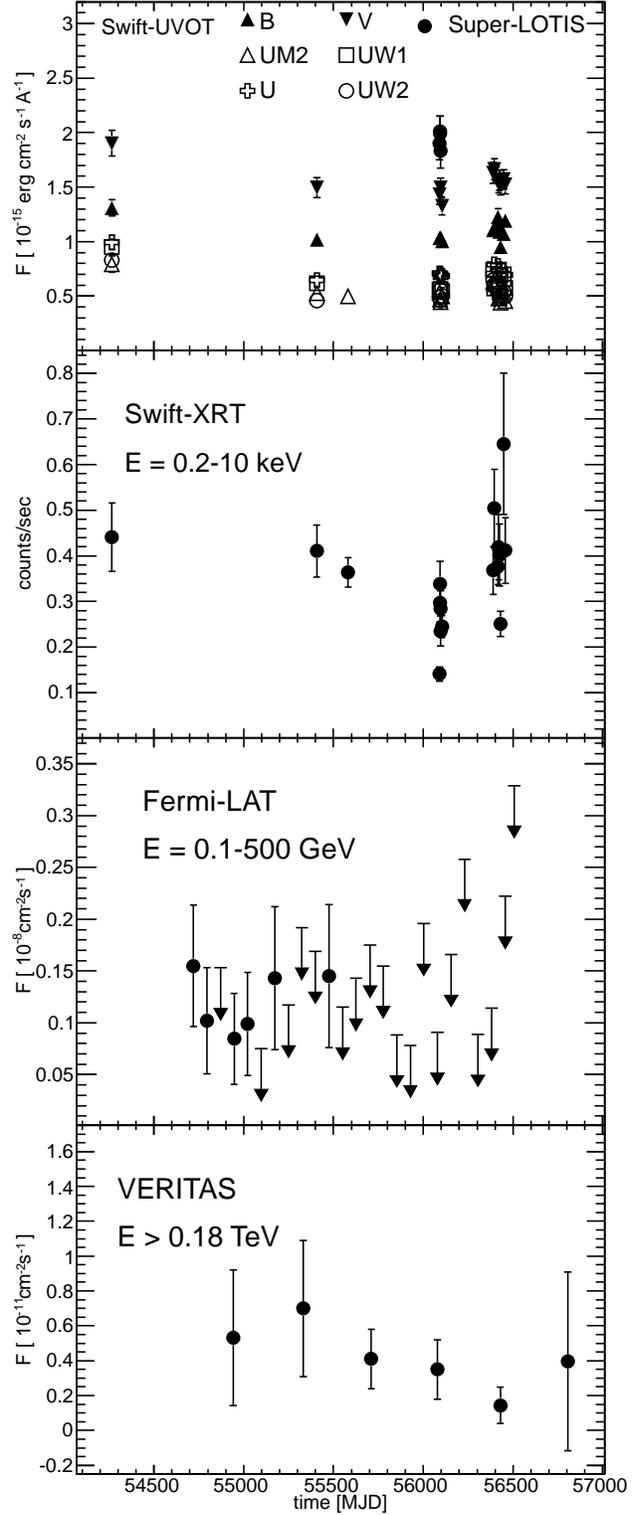}
\caption{Multiwavelength light curve for 1ES 1741+196. From top to bottom, the optical and ultraviolet, X-ray, high-energy, and very-high-energy light curves are shown. The \textit{Swift}--UVOT light curves are derived from data taken with different filters, and exhibit variability for all filters.}
\label{Lightcurve}
\end{center}
 \end{figure}

The flux for each season from 2009 to 2014 is plotted in the bottom panel of Fig.~\ref{Lightcurve}. The light curve is computed from the integral excess counts above an energy threshold of 180 GeV, assuming a power-law spectrum with index of 2.7, as found in the spectral fit to the full data set. The dead-time-corrected exposure times by season are as follows: 2.5 hours in 2009, 2.3 hours in 2010, 9.6 hours in 2011, 5.0 hours in 2012, 8.2 hours in 2013, and 0.6 hours in 2014.

\begin{table}
\caption{VERITAS spectral points, as well as the number of events in the signal ($N_{\mathrm{on}}$) and background ($N_{\mathrm{off}}$) regions in each energy bin. The highest-energy flux is an upper limit at the 95\% confidence level. The ratio $\alpha$ of the areas of the signal and background regions is 0.1.}
\centerline{
\begin{tabular}{ccc}
\hline
Mean Energy [TeV]   &  Flux [m$^{-2}$s$^{-1}$TeV$^{-1}$] & $N_{\mathrm{on}}/N_{\mathrm{off}}$  \\
 \hline
  0.21    &     (2.5 $\pm$ 1.0) $\times$ 10$^{-7}$   &  144/1116    \\
  0.30    &     (9.0 $\pm$ 2.8) $\times$ 10$^{-8}$   &  125/887   \\
  0.42    &     (3.8 $\pm$ 1.2) $\times$ 10$^{-8}$   &  78/527   \\
  0.60    &     $<$ 4.46 $\times$ 10$^{-8}$          &  45/358 \\
\hline
\end{tabular}
}

\label{tab:spectralpts}
\end{table}    
 
\subsection{\textit{Fermi}--LAT}\label{Fermi}
To determine the {\em Fermi}-LAT spectrum of the source, we considered two time intervals: i) all LAT data collected since the start of the science mission in 2008 August until 2014 June (i.e. for 70 months of operation), and ii) time intervals contemporaneous with the VERITAS observations. We define contemporaneous observations as those made on the same date as VERITAS observations. The data were analysed using an official release of the {\em Fermi} ScienceTools (v9r34p1), Pass 7 instrument response functions and considering photons satisfying the SOURCE event selection. We exclude photons detected at instrument zenith angles greater than 100$^{\circ}$~to avoid contamination from the Earth's limb, and data taken with the Earth within the LAT's field of view by requiring a rocking angle less than 52$^{\circ}$. The spectral analysis of each source is based on the maximum likelihood technique using the standard likelihood analysis software. In addition to diffuse background components (Galactic and isotropic) we included all point-like sources within $15^\circ$ of the source position as determined from the 2FGL catalog~\citep{nolan12} and residual Test-Statistic (TS) maps in the maximum likelihood fit.\footnote{Residual hotspots with TS $>$ 10 and less than $10^\circ$ away from the ROI center were included in the model.}

The spectral parameters resulting from our power-law fit to the full 70-month data set are
\begin{multline}
\frac{dN}{dE}  = (4.2 \pm 1.1) \times 10^{-7} \\ \bigg( \frac{E}{4~GeV} \bigg)^{-1.87 \pm 0.10} [m^{-2} s^{-1} MeV^{-1}] 
\end{multline}
where the quoted errors are statistical. The spectral index is consistent within statistical errors with the spectral index of 1.777$\pm$0.108 measured by \cite{3LAC}. The test statistic (TS)~\citep{mattox96} is 103.  Due to the low flux of the source, the contemporaneous data are only sufficient to produce a weak integral upper limit of  $3.7 \times 10^{-5} m^{-2} s^{-1}$ at the 95\% confidence level above an energy threshold of 100 MeV.  The TS is 7.3 for this upper limit.  The light curve is shown in Fig.~\ref{Lightcurve} in time bins of 75 days. Bins with a TS of less than 10 are plotted as upper limits. Significant flux variability is not observed, consistent with the low variability index of 38.3 measured by \cite{3LAC}.

\subsection{\textit{Swift}--XRT}\label{Swift}
The X-ray emission from 1ES 1741+196 has been measured using the X-Ray-Telescope (XRT) on board the \textit{Swift} satellite~\citep{XRT,Swift}. We analysed all observations available on this target, from June 2007 to June 2013, for a total live-time of about 19 ks (see Table~\ref{TableXrays} for details). All observations have been taken using the XRT photon-counting mode. Data are reduced using \textit{HEASoft} (version 6.14). Event files are produced using \textit{Swift} default screening criteria, and images, light curves, and spectra are extracted (using \textit{XSelect}, version 2.4b) from a circular region of radius equal to 20 and 50 pixels for the source and the background, respectively.
   
The full \textit{Swift}--XRT light curve of 1ES 1741+196 is plotted in Fig.~\ref{Lightcurve}. Following the prescriptions of the \textit{Swift} team,\footnote{See \url{http://www.swift.ac.uk/analysis/xrt/pileup.php}} the light curve has been corrected for the exposure and the background, and rebinned to ensure a minimum of 50 counts per bin. The count rates are between 0.1 and 0.6 counts/second, and could potentially be affected by pile-up. This effect has been investigated and it was concluded that no pile-up is present in the observations. There is clear variability in the X-ray data, with a mean fractional variability in the light curve of 27\% (calculated according to the prescription of \cite{Vaughan}). 
   
We studied the X-ray spectrum of 1ES 1741+196 using \textit{XSpec} (version 12.8.0). Given the source variability, we first investigated whether there is any spectral evolution by fitting individual XRT observations with a simple power-law model. Absorption by Galactic material is taken into account using the \textit{tbnew} model, with $N_H = 6.86 \times 10^{20}$ cm$^{-2}$, as provided by~\citet{Dickey90}. We used response functions provided by the \textit{Swift} team, and computed specific ancillary response files using \textit{xrtmkarf}. Data below 0.3 keV are excluded from the analysis and the spectra have been rebinned, imposing a minimum of 30 counts per bin (to perform $\chi^2$ minimization). The results of the power-law fits are provided in Table~\ref{TableXrays}. For two of the observations, the number of spectral bins is too small to perform a fit, and they have been disregarded. For the other 15 observations, the spectra are consistent with power-law emission, the worst fit showing a null hypothesis probability of 3.5\%. 

\begin{figure}
\begin{center}
\includegraphics[width=250pt]{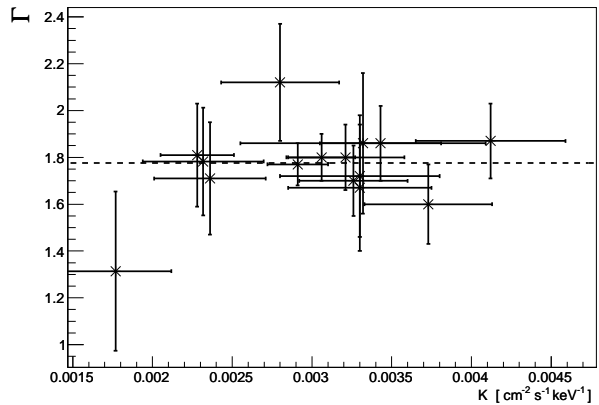}
\caption{Fitted photon index ($\Gamma$) versus normalisation (K) measured in \textit{Swift}--XRT data. The photon indices are consistent with a single value of 1.78, as indicated by the red dashed line, indicating no spectral variability.}
\label{FigXrays}
\end{center}
\end{figure}

Despite the flux variability, the X-ray emission of 1ES 1741+196 does not show significant spectral variability. A fit of the spectral index as a function of the flux is consistent with a constant value ($\chi^2 = 6.2/14$), as shown in Fig.~\ref{FigXrays}.
   
We investigated the average X-ray spectrum by adding all the XRT observations (using \textit{mathpha}). The average spectrum is first fit with an absorbed power-law function, resulting in a photon index $\Gamma=1.76 \pm 0.03$ and normalisation $K=(2.85 \pm 0.07)\times10^{-3}$ cm$^{-2}$ s$^{-1}$ keV$^{-1}$. There are large residuals for the single-component fit ($\chi^2=201/136$), and the fit is significantly improved by using a broken-power-law function that is physically motivated by rapid cooling of the highest energy electrons and slow cooling of the lower energy electrons. The best-fit parameters are $\Gamma_1=1.42 \pm 0.11$, $\Gamma_2=1.99 \pm 0.08$, $E_{break} = 1.20 \pm 0.17$ keV and $K=(3.06 \pm 0.13)\times10^{-3}$ cm$^{-2}$ s$^{-1}$ keV$^{-1}$, with $\chi^2=150/134$. An F-test indicates that the fit with a broken-power-law function is statistically preferred over a simple power-law model, with a null hypothesis probability of $3\times10^{-9}$. We conclude from this that there is curvature in the spectrum and that the peak in the synchrotron emission is above 1.0 keV. In Fig.~\ref{FigSED}, we plot the average XRT spectrum, corrected for Galactic absorption.
     
   \begin{table*}
   	\caption{Summary of \textit{Swift}--XRT observations of 1ES 1741+196}
   	\begin{center}
   		\begin{tabular}{|c|c|c|c|c|c|}
   			\hline
   			Obs. ID & Date & Exposure & Index & K & $\chi^2$/DOF \\
   			& & [ks] & & [$10^{-3}$ cm$^{-2}$ s$^{-1}$ keV$^{-1}$] & \\ 
   			\hline
   			30950001 &  Jun 15, 2007 & 1.9 & $1.80 \pm 0.10$  & $3.06 \pm 0.21$  & 22 / 20  \\
   			40639001 &  Jul 30, 2010 & 0.8 &$1.67 \pm 0.27$  &  $3.30 \pm 0.45$  & 4 / 6  \\   
   			40639002 &  Jan 21, 2011& 2.9 &$1.77 \pm 0.09$  &  $2.91 \pm 0.19$ &   20 / 25\\
   			40639003 &  Jun 13, 2012& 0.9 &- &  - & -  \\
   			40639004  & Jun 15, 2012& 1.1  &$1.71 \pm 0.24$  & $2.36 \pm 0.35$  &   2 / 5\\
   			40639005   &Jun 16, 2012& 1.0 &$2.12 \pm 0.25$  & $2.80 \pm 0.37$  &   3 / 5\\
   			40639006   &Jun 18, 2012& 0.9 &$1.78 \pm 0.23$  & $2.32 \pm 0.38$  &   4 / 4\\
   			40639007   &Jun 19, 2012& 1.1 &$1.81 \pm 0.22$  & $2.28 \pm 0.23$  &   8 / 3\\
   			40639008   &Jun 26, 2012&1.0 &$1.31 \pm 0.34$   & $1.77 \pm 0.35$  &   1 / 3\\
   			40639009   &Apr 08,  2013& 0.9 &$1.72 \pm 0.26$   & $3.30 \pm 0.50$  &   3 / 5\\
   			40639010   &Apr 14, 2013& 0.9 &$1.60 \pm 0.17$  &$3.73 \pm 0.40$   &   12 / 10\\
   			40639012   &May 05, 2013& 0.3 &-  & -  & -  \\
   			40639013   &May 08, 2013& 1.1 &$1.70 \pm 0.15$  & $3.26 \pm 0.34$  &   10 / 10\\
   			40639014   &May 12, 2013& 1.0 &$1.86 \pm 0.16$  & $3.43 \pm 0.38$  &   8 / 8\\
   			40639015   &May 19, 2013&0.9 &$1.86 \pm 0.30$  & $3.32 \pm 0.77$  &   4 / 2\\
   			40639016   &Jun 05, 2013&0.9&$1.87 \pm 0.16$   &$4.12 \pm 0.47$   &   9 / 8\\
   			40639018   &Jun 15, 2013&1.0 &$1.80 \pm 0.14$  & $3.21 \pm 0.37$  &   19 / 9\\
   			\hline
   		\end{tabular}
   	\end{center}
   	\label{TableXrays}
   \end{table*}%
   
\subsection{\textit{Swift}--UVOT and Super-LOTIS}
The UVOT telescope~\citep{UVOT} on board the \textit{Swift} satellite observes the same field as the XRT but in the optical and ultraviolet bands, providing simultaneous MWL observations. Six filters are available for UVOT observations: V and B in visible light, and U, UW1, UW2 and UM2 in ultraviolet. In general, for every XRT observation of 1ES 1741+196 there are UVOT images for every filter, however, in a few cases only a few filters were used (Observations 40639002/4/5/6/12). The data analysis was performed using \textit{uvotmaghist}, estimating the source (background) flux from a circular region of 5$\arcsec$ (15$\arcsec$) radius. Data have been corrected for galactic absorption using $E_{B-V}$=0.1~\citep[a value in agreement with the $N_H$ value used in the X-ray analysis, following][]{Jenkins74}.
   
Optical observations of 1ES 1741+196 were also taken using the Super-LOTIS telescope (located at the Steward Observatory Kitt Peak site) with the R filter~\citep{SuperLOTIS}. Observations were performed during 2012 June 13-15-16-18, quasi-simultaneously (i.e. on the same nights) with \textit{Swift}. 
   
In Fig.~\ref{Lightcurve} we show the UVOT and Super-LOTIS light-curves: 1ES 1741+196 is variable in both optical and ultra-violet. In Fig.~\ref{FigSED} we show the average measurements. Because the variability range in every filter is larger than the statistical errors, we have used the absolute variability range for the error bars on the SED.
 
 As discussed in the introduction, the host galaxy of 1ES 1741+196 is an interacting giant elliptical, and this complicates the subtraction of the host contribution. Although~\cite{Heidt} provide the flux and the effective radius of the host, only the R filter is used. It is not possible to correctly estimate the host contribution in the other filters. The galaxy profile is not de Vaucouleurs-like, and star-forming regions may cause additional contamination. In Fig.~\ref{FigSED}, we present the optical-UV data without attempting any host-galaxy subtraction. Consequently, the optical-UV flux points in Fig.~\ref{FigSED} should be regarded as upper limits.

\section{SED Modelling}\label{SED}
\begin{figure}
 \begin{center}
 \includegraphics[width=250pt]{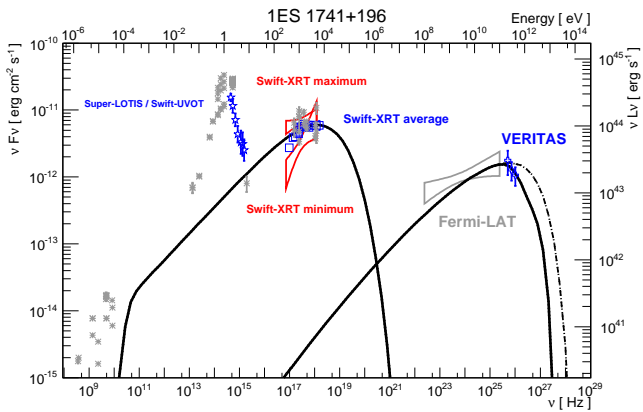}
 \caption{Multiwavelength spectral energy distribution for 1ES 1741+196. Grey symbols are archival data, blue symbols are observations analysed here: VERITAS (open cross), \textit{Swift}--XRT (open square), Super-LOTIS and \textit{Swift}--UVOT (open stars).  The maximum and minimum \textit{Swift}--XRT bow-ties are also shown (red) as well as the average \textit{Fermi}--LAT bow-tie (grey).  Note that the lowest frequency peak, the thermal component due to the host galaxy, is not included in the non-thermal SED model. A single SSC model is overlaid on the SED, showing consistency between the data and a one-zone SSC scenario. The same model, including EBL absorption (using the model of Franceschini 2008), is also shown by the dashed line.}
 \label{FigSED}
 \end{center}
 \end{figure}

The spectral energy distribution of 1ES 1741+196 is shown in Fig.~\ref{FigSED}. It includes the VERITAS and multiwavelength data presented in this work, as well as archival measurements.  The SED is typical of high-frequency-peaked blazars, with two broad non-thermal components peaking in X-rays and $\gamma$-rays, respectively.  In addition, there is a low energy thermal peak that is attributed to the host galaxy which is not modelled.  In the VHE regime, the model is shown with and without EBL absorption. The model including EBL absorption utilizes the model of~\cite{Franceschini}. 

In the framework of the synchrotron-self-Compton (SSC) model, the first non-thermal component is due to synchrotron emission by leptons (electrons and positrons) in a blob of plasma in the relativistic jet, while the high-energy non-thermal component is due to inverse-Compton scattering between the leptons and their own synchrotron photon field. The free parameters of the model are the Doppler factor $\delta$ and the radius $R$ of the emitting region (assumed spherical), the magnetic field $B$, and the parameters of the lepton energy distribution, which is assumed to be parametrized by a  simple power-law distribution, characterized by the index $\alpha$, the normalisation $K$ (particle density in units of cm$^{-3}$) and the minimum and maximum Lorentz factors $\gamma_{min}$ and $\gamma_{max}$. The model used is the one described in~\cite{Cerruti13}. 

The modelling of the SED is complicated by the fact that there is no simultaneous \textit{Fermi}--LAT detection, and the $\gamma$-ray emission is measured only as an average flux state integrated over several observing seasons (while there is clear variability in the synchrotron component). In addition, the unusual host galaxy makes it difficult to extract the underlying non-thermal continuum in the optical/UV part of the spectrum. For these reasons we do not perform a fit of the SED.  We present instead a single SSC model as an example, showing that the average emission is compatible with a one-zone SSC scenario, as for other HBLs detected by Cherenkov telescopes. We searched for SSC solutions which correctly describe the VERITAS data and the average \textit{Swift}--XRT spectrum. We also used the \textit{Swift}--UVOT and Super-LOTIS data as upper limits for the blazar emission. Although we do not explicitly model the average \textit{Fermi}--LAT spectrum, the model is compatible with its spectral shape and underestimates the overall normalisation by only a factor of two. The average SED is well described assuming a Doppler factor $\delta=20$, a magnetic field $B=40\ \textrm{mG}$, a radius $R=1\times10^{16}\ \textrm{cm}$, and a population of leptons with $\gamma$ between $\gamma_{min}=10$ and $\gamma_{max}=1\times10^6$, with index $\alpha=2.2$ and normalisation $K=8000$  cm$^{-3}$. The parameters of the SSC model are in agreement with the ones used for other VHE blazars~\citep[see e.g.][]{Tavecchio10, Zhang14}. 

The model presented here describes the average \textit{Swift}--XRT emission, but can easily be fine-tuned to describe instead the minimum and maximum soft-X-ray spectra. This is accomplished by varying the normalisation $K$ and the emitting region radius $R$, which changes the ratio between the synchrotron and the inverse-Compton components.
     
\section{Discussion}\label{discussion}    
One of the most discussed topics in blazar physics is the so-called \textit{blazar sequence}, an anti-correlation between the blazar synchrotron peak-frequency and luminosity~\citep{Fossati98}. In this scenario, HBLs are the lowest luminosity blazars, but with the highest peak frequencies. A natural question arises: is the blazar sequence truncated at its low end, or are there even dimmer and higher-frequency-peaked blazars? In recent years, observations with Cherenkov telescopes have led to the discovery of extreme-HBLs, also called ultra-HBLs~\cite[see][and references therein]{Costamante01, Bonnoli15, Cerruti15}. The number of detected extreme-HBLs is not large. The BL Lac object 1ES 0229+200 can be considered the best example of this peculiar blazar sub-class, with a synchrotron peak-frequency at more than 10 keV (i.e. out of the range of soft-X-ray telescopes) and an intrinsic inverse-Compton peak-frequency at several TeV~\citep[see][]{0229hess, 0229veritas}. The source can be described with a one-zone SSC model~\citep{0229veritas}. Other extreme-HBLs detected at VHE are 1ES 0347-121~\citep{0347hess}, RGB J0710+591~\citep{0710veritas}, 1ES 1101-232~\citep{1101hess} and 1ES 1218+304~\citep{1218magic,1218veritas}. 

The SED of 1ES 1741+196 is particularly interesting because it suggests that this VHE blazar may belong to the extreme-HBL class. The synchrotron peak-frequency is constrained by the average \textit{Swift}--XRT spectrum, which is best-fit by a broken-power-law with $\Gamma_2=1.99 \pm 0.08$, $E_{break} = 1.20 \pm 0.17$ keV, indicating a synchrotron peak-frequency located above 1.0 keV. At the same time, the VERITAS spectrum corrected for EBL absorption is consistent with $\Gamma_{VHE}\sim 2.3$, indicating an inverse-Compton peak-frequency above 100 GeV. Further observations are needed to determine the peak locations precisely.  Although the peak-frequency may not be as extreme as 1ES~0229+200, these lower limits are elevated compared to other well-studied HBLs in their non-flaring states, such as Mrk 421~\citep{Mrk421veritas}, Mrk 501~\citep{Mrk501veritas}, or PKS 2155-304~\citep{2155hess}.

A further point of interest for this source is the lack of strong flares in the VHE band. The statistical uncertainties on the yearly flux points exceed the variation on the central values. Variability is therefore not ruled out, but strong flaring behaviour is not observed. The majority of blazars show substantial variability over yearly time scales or shorter in the HE and VHE bands. The detection of a blazar via a multi-year observing campaign, rather than a short period of observations during a flare, marks 1ES 1741+196 as an unusual source. Interestingly, three extreme-HBLs, 1ES 0347-121, RGB J0710+591, and 1ES 1101-232 are notable for their lack of detected variability in the VHE band. A deeper observing campaign would be necessary to determine whether 1ES 1741+196 exhibits the same behaviour.

Extreme-HBLs have also attracted interest because their spectra extend to high energies, making them particularly good probes for constraining the EBL~\citep[see e.g.][]{0229hess} and the intergalactic magnetic field (IGMF)~\citep[see e.g.][]{Dermer11}.  With contemporaneous measurements of the inverse-Compton regime, 1ES 1741+196 would be an interesting candidate for EBL studies. However, due to the low flux of the source, accumulating enough $\gamma$-ray events to make a precise measurement of the spectrum for EBL constraints would require either a long observation window or observations during a period of elevated flux. This is highlighted by the $Fermi$-LAT light curve, which shows non-detections for many 75-day exposures. As a VERITAS measurement of the differential flux requires $\sim$30 hours, for useful contemporaneous $\gamma$-ray observations, VERITAS would need to collect on the order of 60 hours of observations during the single six month period that is likely to yield a $Fermi$-LAT detection. These challenging observing campaigns will be simpler for future instruments that are more sensitive, such as the Cherenkov Telescope Array (CTA). 

\section{Conclusions}\label{conclusions}
We have presented the first VHE spectrum and MWL modelling of the SED for the BL Lac 1ES 1741+196. The blazar has a relatively hard spectrum, and the MWL SED is consistent with a simple one-zone SSC model. The source is relatively dim, at $\sim$1.6\% of the Crab Nebula flux above 180 GeV, which makes simultaneous multiwavelength observations especially challenging. It is also located within a group of interacting galaxies, which makes the analysis of the host-galaxy emissions difficult. Despite these difficulties, several uncommon characteristics make this blazar an interesting target. The synchrotron and inverse-Compton emissions are peaked at very high energies for an HBL, pointing to classification as an extreme-HBL. Additionally, the source does not show evidence for strong flares in any energy range. The source is potentially interesting for EBL studies and other cosmological measurements, but long exposures would be necessary for such studies, in light of the source's low $\gamma$-ray flux.

\section*{Acknowledgments}
This research is supported by grants from the U.S. Department of Energy Office of Science, the U.S. National Science Foundation and the Smithsonian Institution, and by NSERC in Canada. This research used computational resources of the National Energy Research Scientific Computing Center, a DOE Office of Science User Facility supported by the Office of Science of the U.S. Department of Energy under Contract No. DE-AC02-05CH11231. E. Pueschel acknowledges the support of a Marie Curie Intra-European Fellowship within the 7th European Community Framework Programme. We acknowledge the excellent work of the technical support staff at the Fred Lawrence Whipple Observatory and at the collaborating institutions in the construction and operation of the instrument. We are also grateful to Grant Williams and Daniel Kiminki for their dedication to the operation and support of the Super-LOTIS telescope. The VERITAS Collaboration is grateful to Trevor Weekes for his seminal contributions and leadership in the field of VHE $\gamma$-ray astrophysics, which made this study possible.




\bsp

\label{lastpage}

\end{document}